\documentstyle[12pt]{article}
\begin{document}
\begin{flushright}
arXiv:0704.0064 [hep-th]\\ CAS-PHYS-BHU/Preprint
\end{flushright}
\vskip 2.5cm
\begin{center}
{\bf NILPOTENT SYMMETRY INVARIANCE IN THE SUPERFIELD
FORMULATION: THE (NON-)ABELIAN 1-FORM GAUGE THEORIES}

\vskip 2.5cm

  R. P. MALIK\\
{\it Centre of Advanced Studies, Physics Department,}
\\ {\it Banaras Hindu University, Varanasi- 221 005, (U. P.), India}
\\ {\small  E-mails: rudra.prakash@hotmail.com ; malik@bhu.ac.in}\\

\vskip 2cm

\end{center}

\noindent {\bf Abstract}: We capture the off-shell as well 
as the on-shell nilpotent Becchi-Rouet-Stora-Tyutin (BRST)
and anti-BRST symmetry invariance of the Lagrangian densities
of the four (3 + 1)-dimensional (4D) (non-)Abelian 1-form gauge
theories within the framework of the superfield formalism. In
particular, we provide the geometrical interpretations for (i) the
above nilpotent symmetry invariance, and (ii) the above
Lagrangian densities, in the language of the specific quantities
defined in the domain of the above superfield formalism.
Some of the subtle points, connected with the 4D (non-)Abelian
1-form gauge theories, are clarified within the framework of the
above superfield
formalism where the 4D ordinary gauge theories are considered on
the (4, 2)-dimensional supermanifold parametrized by the four
spacetime coordinates $x^\mu$ (with $\mu = 0, 1, 2, 3$) and a pair
of Grassmannian variables $\theta$ and $\bar\theta$. One of the
key results of our present investigation is a great deal of
simplification in the geometrical understanding of the nilpotent 
(anti-)BRST symmetry invariance. 
\\

\baselineskip=16pt

\vskip .7cm



\newpage

\noindent {\bf 1 Introduction}\\

\noindent The geometrical superfield approach [1-8] to
Becchi-Rouet-Stora-Tyutin (BRST) formalism is one of the most
attractive and intuitive approaches which enables us 
to gain some physical insights
into the beautiful (but abstract mathematical) structures that are
associated with the nilpotent (anti-)BRST symmetry transformations
and their corresponding generators. The latter quantities play a
very decisive role in (i) the covariant canonical quantization of
the gauge theories, (ii) the proof of the unitarity of the
``quantum'' gauge theories at any arbitrary order of perturbative
computations for a given physical process (that is allowed by
the theory), (iii) the definition of
the physical states of the ``quantum'' gauge theories in the
quantum Hilbert space, and (iv) the
cohomological description of the physical states of the quantum
Hilbert space w.r.t. the conserved and nilpotent BRST charge.

To be specific, in the superfield formulation [1-8] of the 4D
1-form gauge theories, one defines the super curvature 2-form
$\tilde F^{(2)} = \tilde d \tilde A^{(1)} + i \; \tilde A^{(1)}
\wedge \tilde A^{(1)}$ in terms of the super exterior derivative
$\tilde d = dx^\mu
\partial_\mu + d \theta
\partial_\theta + d \bar\theta
\partial_{\bar\theta}$ (with $\tilde d^2 = 0$) and the super 1-form
connection $\tilde A^{(1)}$ on a (4, 2)-dimensional supermanifold
parametrized by the usual spacetime variables $x^\mu$ (with $\mu =
0, 1, 2, 3$) and a pair of anticommuting (i.e. $\theta^2 =
\bar\theta^2 = 0, \theta \bar\theta + \bar\theta \theta = 0$)
Grassmannian variables $\theta $ and $\bar\theta$. The above super
2-form is subsequently equated, due to the so-called horizontality
condition [1-8], to the ordinary curvature 2-form $F^{(2)} = d
A^{(1)} + i A^{(1)} \wedge A^{(1)}$ defined on the ordinary 4D
flat Minkowski spacetime manifold in terms of the ordinary
exterior derivative $d = dx^\mu \partial_\mu$ (with $d^2 = 0$) and
the 1-form connection $A^{(1)} = dx^\mu A_\mu$. The above super
exterior derivative $\tilde d$ and super 1-form connection $\tilde
A^{(1)}$ are the generalization of the 4D ordinary exterior
derivative $d$ and 1-form connection $A^{(1)}$ to the (4,
2)-dimensional supermanifold because $\tilde d \to d, \tilde
A^{(1)} \to A^{(1)}$ in the limit $(\theta, \bar\theta) \to 0$.

The above horizontality condition (HC) has been referred to as the
soul-flatness condition in [9] which amounts to setting equal to
zero all the Grassmannian components of the (anti)symmetric
second-rank super tensor that constitutes the super curvature
2-form $\tilde F^{(2)}$ on the (4, 2)-dimensional supermanifold.
The key consequences, that emerge from the HC, are (i) the
derivation of the nilpotent (anti-)BRST symmetry transformations
for the gauge and (anti-)ghost fields of a given 4D 1-form gauge
theory, (ii) the geometrical interpretation of the (anti-)BRST
symmetry transformations for the 4D local fields as the
translation of the corresponding superfields along the
Grassmannian directions of the  supermanifold, (iii) the
geometrical interpretation of the nilpotency property as a pair of
successive translations of the superfield along a particular
Grassmannian direction of the supermanifold, and (iv) the
geometrical interpretation of the anticommutativity property of
the (anti-)BRST symmetry transformations for a 4D local field as
the {\it sum} of (a) the translation of the corresponding superfield
first along the $\theta$-direction followed by the translation
along the $\bar\theta$-direction, and (b) the translation of the
same superfield first along the $\bar\theta$-direction followed by
the translation along the $\theta$-direction.

It will be noted that the above HC (i.e. $\tilde F^{(2)} =
F^{(2)}$) is valid for the non-Abelian (i.e. $A^{(1)(n)}
\wedge  A^{(1)(n)} \neq 0$) 1-form gauge theory as well as the Abelian (i.e.
$ A^{(1)} \wedge A^{(1)} = 0$) 1-form gauge theory. As expected, for both
types of theories, the HC leads to the derivation of the nilpotent
(anti-)BRST symmetry transformations for the gauge and
(anti-)ghost fields of the respective theories. We lay emphasis on
the fact that the HC does not shed any light on the derivation of
the nilpotent (anti-)BRST symmetry transformations associated with
the {\it matter} fields of the {\it interacting} 4D (non-)Abelian
1-form gauge theories.

In a recent set of papers [10-17], the above HC condition has been
generalized, in a consistent manner, so as to compute the nilpotent
(anti-)BRST symmetry transformations associated with the {\it
matter} fields of a given 4D interacting 1-form gauge theory
(along with the well-known nilpotent transformations for the gauge and
(anti-)ghost fields) without spoiling the cute geometrical
interpretations of the (anti-)BRST symmetry transformations (and
their corresponding generators) that emerge from the HC {\it
alone}. The latter approach has been christened as the augmented
superfield approach to BRST formalism where the restrictions
imposed on the (4, 2)-dimensional superfields are (i) the HC plus
the invariance of the (super) matter Noether conserved currents
[10-14], (ii) the HC plus the equality of any (super) conserved
quantities [15], (iii) the HC plus a restriction that owes its
origin to the gauge invariance and the (super) covariant derivatives on the matter
(super)fields [16,17], and (iv) an alternative to the HC where the gauge
invariance and the property of a pair of (super) covariant derivatives
on the (super) matter fields (and their intimate connection with the (super) 
curvatures) play a crucial role [18-20].

In all the above approaches [1-20], however, the invariance of the
Lagrangian densities of the 4D (non-)Abelian 1-form gauge theories,
under the nilpotent (anti-)BRST symmetry transformations, has not
yet been discussed at all. Some attempts in this direction have
been made in our earlier works where the specific topological
features [21,22] of the 2D free (non-)Abelian 1-form gauge theories have
been captured in the superfield formulation [23-25]. In particular, the
invariance of the Lagrangian density under the nilpotent and anticommuting
(anti-)BRST and (anti-)co-BRST symmetry transformations has been
expressed in terms of the superfields and the Grassmannian
derivatives on them. These are, however, a bit more involved in
nature because of the existence of a new set of nilpotent
(anti-)co-BRST symmetries in the theory. The geometrical
interpretations for the Lagrangian densities and the symmetric
energy-momentum tensor (for the above topological theory) have
also been provided within the framework of the superfield formulation.

The purpose
of our present paper is to capture the (anti-)BRST symmetry
invariance of the Lagrangian density of the 4D 
(non-)Abelian 1-form gauge theories within the framework of the
superfield approach to BRST formalism and to demonstrate that the
above symmetry invariance could be understood in a very simple
manner in terms of the translational generators along the
Grassmannian directions of the (4, 2)-dimensional supermanifold
on which the above 4D ordinary gauge theories are considered. In addition,
the reason behind the existence (or non-existence) of any specific
nilpotent symmetry transformation could also be explained within
the framework of the above superfield approach.
We demonstrate the {\it uniqueness}
of the existence of the 
nilpotent (anti-)BRST symmetry transformations for the Lagrangian
density of a U(1) Abelian 1-form gauge theory. We go a step
further and show the existence of the 
nilpotent BRST symmetry transformations for the specific
Lagrangian densities (cf. (4.1) and (4.4) below) of the 4D
non-Abelian 1-form gauge theory and clarify the non-existence of the
anti-BRST symmetry transformations for these specific Lagrangian
densities within the framework of the superfield formulation (cf.
section 5 below). Finally, we provide the geometrical basis for
the existence of the off-shell nilpotent and anticommuting (anti-)BRST symmetry
transformations (and their corresponding generators) 
for the specifically defined Lagrangian densities
(cf. (4.7) and/or (4.8) below) of the 4D non-Abelian 1-form gauge
theory in the Feynman gauge.

The motivating factors that have propelled us to pursue our
present investigation are as follows. First and foremost, to the
best of our knowledge, the property of the symmetry invariance of
a given Lagrangian density has not yet been captured in the
language of the superfield approach to BRST formalism. Second, the
above (anti-)BRST invariance of the theory has never been shown,
in as simplified fashion, as we demonstrate in our present
endeavour. The geometrical interpretations for (i) the
existence of the above nilpotent (anti-)BRST symmetry
invariance, and (ii) the on-shell conditions of
the on-shell nilpotent (anti-)BRST symmetries,
turn out to be quite transparent in our present work. 
Third, we establish
the {\it uniqueness} of the existence of the (anti-)BRST symmetry
invariance in their various forms. The non-existence of the
specific symmetry transformation is also explained within the
framework of the superfield approach to BRST formalism. Finally,
our present investigation is the first modest step in the
direction to gain some insights into the existence of the
nilpotent symmetry transformations and their invariance 
for the higher form (e.g.
2-form, 3-form, etc.) gauge theories within the framework of the 
superfield formulation.

The contents of our present paper are organized as follows.
In section 2, we recapitulate some of the key points connected
with the nilpotent (anti-)BRST
symmetry transformations for the free 4D Abelian 1-form gauge theory
(having no interaction with matter fields) in the Lagrangian
formulation. The above symmetry transformations as well as the symmetry
invariance of the Lagrangian densities are captured in the
geometrical superfield approach to BRST formalism in section 3
where the HC on the gauge superfield 
plays a crucial role. Section 4 deals with the bare essentials of 
the nilpotent (anti-)BRST symmetry transformations
for the 4D non-Abelian 1-form gauge theory in the Lagrangian formulation.
The subject matter of section 5
concerns itself with the superfield formulation of the
symmetry invariance of the appropriate Lagrangian densities of
the above 4D non-Abelian 1-form gauge theory.
Finally, in section 6, we summarize our key results, make some
concluding remarks and point out a few future directions for
further investigations.\\

\noindent {\bf 2 (Anti-)BRST symmetries in Abelian theory:
Lagrangian formulation}\\

\noindent Let us begin with the following (anti-)BRST invariant
Lagrangian density of the 4D Abelian 1-form gauge
theory\footnote{We adopt here the notations and conventions such
that the flat Minkowski metric in 4D is $\eta_{\mu\nu} =$ diag
$(+1, -1, -1, -1)$ so that $A_\mu B^\mu = \eta_{\mu\nu} A^\mu
B^\nu = A_0 B_0 - A_i B_i$ for two non-null 4-vectors $A_\mu$ and
$B_\mu$. The Greek indices $\mu, \nu......= 0, 1, 2, 3$ and Latin
indices $i, j, k....= 1, 2, 3$ stand for the 4D spacetime and 3D space
directions on the 4D Minkowski spacetime manifold, respectively, and 
the symbol $\Box
= (\partial_0)^2 - (\partial_i)^2$.}
in the Feynman gauge
[26,27,9] $$
\begin{array}{lcl}
{\cal L}^{(a)}_B = {\displaystyle - \frac{1}{4}\; F^{\mu\nu}
F_{\mu\nu} \;+ \;B\; (\partial_\mu A^\mu)\; +\; \frac{1}{2}\; B^2
- i\;\partial_\mu \bar C\;\partial^\mu C},
\end{array} \eqno(2.1)
$$ where $F_{\mu\nu} = \partial_\mu A_\nu - \partial_\nu A_\mu$ is
the antisymmetric ($F_{\mu\nu} = - F_{\nu\mu}$) curvature tensor
that constitutes the Abelian 2-form $F^{(2)} = d A^{(1)} \equiv
\frac{1}{2!} (dx^\mu \wedge dx^\nu) F_{\mu\nu}$, $B$ is the
Nakanishi-Lautrup auxiliary multiplier field and $(\bar C)C$ are
the anticommuting (i.e. $C^2 = \bar C^2 = 0, C \bar C + \bar C C =
0$) (anti-)ghost fields of the 
theory. The above Lagrangian
density respects the off-shell nilpotent ($s_{(a)b}^2 = 0$)
(anti-)BRST symmetry transformations $s_{(a)b}$ (with $s_b s_{ab}
+ s_{ab} s_b = 0$)\footnote{We follow here the notations and
conventions adopted in [27]. In its full blaze of glory, the
nilpotent (anti-)BRST transformations $\delta_{(A)B}$ are a
product of an anticommuting spacetime independent parameter $\eta$
and $s_{(a)b}$ (i.e. $\delta_{(A)B} = \eta s_{(a)b}$) where the
nilpotency property is encoded in the operators $s_{(a)b}$.} $$
\begin{array}{lcl}
&& s_b A_\mu = \partial_\mu C,\; \qquad s_b C = 0,\; \qquad s_b \bar C
= i B,\; \qquad s_b B = 0,\; \qquad s_b F_{\mu\nu} = 0,\; \nonumber\\ &&
s_{ab} A_\mu =
\partial_\mu \bar C,\; \quad s_{ab} \bar C = 0,\; \;\quad s_{ab} C = - i
B,\; \quad s_{ab} B = 0,\; \qquad s_{ab} F_{\mu\nu} = 0.
 \end{array} \eqno(2.2)
$$ It is clear that, under the nilpotent (anti-)BRST symmetry
transformations $s_{(a)b}$, the curvature tensor $F_{\mu\nu}$ is
found to be invariant. In other words, the 2-form $F^{(2)}$, owing
its origin to the cohomological operator $d = dx^\mu
\partial_\mu$, is an (anti-)BRST invariant object for the
Abelian U(1) 1-form gauge theory and is, therefore, 
a physically meaningful
(i.e. gauge-invariant) quantity. These observations will play an
important role in our discussion on the horizontality condition
that would be exploited in the context of our superfield approach
to (anti-)BRST invariance of the Lagrangian densities in
sections 3 and 5 (see below).

A noteworthy point, at this stage, is the observation that the
gauge-fixing and Faddeev-Popov ghost terms can be written, modulo
a total derivative, in the following fashion $$
\begin{array}{lcl}
&& {\displaystyle s_b \Bigl [\;- i \;\bar C \;\{ (\partial_\mu
A^\mu) + \frac{1}{2}\; B\}]}, \;\;\;\qquad\;\;\;
 {\displaystyle s_{ab} \Bigl [\;+ i\;  C\; \{ (\partial_\mu
A^\mu) + \frac{1}{2}\; B\} \; \Bigr ]}, \nonumber\\
&& {\displaystyle s_b\; s_{ab} \;\Bigl [\; \frac{i}{2}\; A_\mu\;
A^\mu + \frac{1}{2}\; C \;\bar C\; \Bigr ]}.
\end{array} \eqno(2.3)
$$ The above equation establishes, in a very simple manner, the
(anti-)BRST invariance of the 4D Lagrangian density (2.1). The
simplicity ensues due to (i) the
nilpotency $s_{(a)b}^2 = 0$ of the (anti-)BRST symmetry
transformations, (ii) the anticommutativity property (i.e. $s_b
s_{ab} + s_{ab} s_b = 0$) of $s_{(a)b}$, and (iii) the invariance
of the $F_{\mu\nu}$ term under $s_{(a)b}$.

As a side remark, it is interesting to note that the following 
on-shell (i.e. $\Box C = \Box \bar C = 0$) nilpotent 
($\tilde s_{(a)b}^2 = 0$) (anti-)BRST symmetry transformations
(with $\tilde s_b \tilde s_{ab} + \tilde s_{ab} \tilde s_b = 0$)
$$
\begin{array}{lcl}
&& \tilde s_b A_\mu = \partial_\mu C,\; \qquad \;\tilde s_b C = 0,\; \qquad \;
\tilde s_b \bar C
= - i (\partial_\mu A^\mu),\; \qquad \;\tilde s_b F_{\mu\nu} = 0,\; \nonumber\\ &&
\tilde s_{ab} A_\mu =
\partial_\mu \bar C,\; \qquad \tilde s_{ab} \bar C = 0,\; \;\qquad \tilde 
s_{ab} C = + i (\partial_\mu A^\mu),\; \qquad  \tilde s_{ab} F_{\mu\nu} = 0,
 \end{array} \eqno(2.4)
$$  
are the symmetry transformations for the following Lagrangian density
$$
\begin{array}{lcl}
{\cal L}^{(a)}_b = {\displaystyle - \frac{1}{4}\; F^{\mu\nu}
F_{\mu\nu} \; -\; \frac{1}{2}\;(\partial_\mu A^\mu)^2\; 
- i\;\partial_\mu \bar C\;\partial^\mu C}.
\end{array} \eqno(2.5)
$$ 
The above transformations (2.4) and the Lagrangian density (2.5) have been
derived from (2.2) and (2.1) by the substitution $B = - (\partial_\mu A^\mu)$.
An interesting point, connected with the on-shell nilpotent symmetry transformations,
is to express the analogue of (2.3) as \footnote{We lay emphasis on the fact that
(2.6) {\it cannot} be derived directly from  (2.3) by the
simple substitution $B = - (\partial_\mu A^\mu)$. One has to be judicious
to deduce the precise expression for (2.6). The logical reasons
behind the derivation of (2.6) are encoded in the superfield formulation 
(cf. (3.9) below).}
$$
\begin{array}{lcl}
&& {\displaystyle \tilde s_b \Bigl [\;+ \frac{i}{2} \;\bar C \;(\partial_\mu
A^\mu) + i\;A_\mu \partial^\mu \bar C ]}, \;\;\;\qquad\;\;\;
 {\displaystyle \tilde s_{ab} \Bigl [\;- \frac{i}{2}\;  C\;  (\partial_\mu
A^\mu) - i\; A_\mu \partial^\mu C \; \Bigr ]}, \nonumber\\
&& {\displaystyle \tilde s_b\; \tilde s_{ab} \;\Bigl [\; \frac{i}{2}\; A_\mu\;
A^\mu + \frac{1}{2}\; C \;\bar C\; \Bigr ]}.
\end{array} \eqno(2.6)
$$ 
It should be noted that, in the above precise computation, one has to take
into account the on-shell ($\Box C = \Box \bar C = 0$) conditions so that, for
all practical purposes $\tilde s_{(a)b} (\partial_\mu A^\mu) = 0$.

The above nilpotent (anti-)BRST symmetry transformations (i.e.
$s_r, \tilde s_r$ with $r = b, ab$) are connected with the
conserved and nilpotent generators (i.e. $Q_r, \tilde Q_r$ with $r
= b, ab$). This statement can be succinctly expressed, in the
mathematical form, as $$
\begin{array}{lcl}
s_r\; \Omega = - i\; [\; \Omega,\; Q_r \;]_{(\pm)},\; \quad
\tilde s_r\; \tilde \Omega = - i\; [\; \tilde \Omega,\; \tilde Q_r \;]_{(\pm)},\; 
\qquad \;\;r = b, ab,
\end{array}\eqno(2.7)
$$ where the subscripts (with the signatures $(\pm)$) on the square
bracket stand for the bracket to be an (anti)commutator, for the
generic fields $\Omega = A_\mu, C, \bar C, B$ 
and $\tilde \Omega = A_\mu, C, \bar C$ 
(of the Lagrangian densities (2.1) and (2.5)),
being (fermionic)bosonic in nature. The above 
charges $Q_r, \tilde Q_r$ are found to be 
anticommuting (i.e. $Q_b Q_{ab} + Q_{ab} Q_b = 0, \tilde Q_b
\tilde Q_{ab} + \tilde Q_{ab} \tilde Q_b = 0$) and off-shell as well
as on-shell nilpotent ($Q_{(a)b}^2 = 0, \tilde Q_{(a)b}^2 = 0)$
in nature, respectively.\\

\noindent {\bf 3 (Anti-)BRST invariance in Abelian theory:
superfield formalism}\\

\noindent In this section, we exploit the geometrical superfield
approach to BRST formalism, endowed with the theoretical arsenal
of the horizontality condition, to express the (anti-)BRST
symmetry transformations and the Lagrangian densities (cf. (2.1)
and (2.5)) in terms of the superfields defined on the (4,
2)-dimensional supermanifold. The latter is parametrized by the
spacetime coordinates $x^\mu$ (with $\mu = 0, 1, 2, 3$) and a pair
of Grassmannian variables $\theta$ and $\bar\theta$. As a consequence,
the generalization
of the 4D ordinary exterior derivative $d = dx^\mu \partial_\mu$
and the 1-form connection $A^{(1)} = dx^\mu A_\mu (x)$ on the (4,
2)-dimensional supermanifold, are 
$$
\begin{array}{lcl}
&& d \to \tilde d = dx^\mu \;\partial_\mu \;+ \;d \theta\;
\partial_\theta \; + \;d \bar\theta \;\partial_{\bar\theta},\;\;\; \qquad \; \;
\tilde d^2 = 0, \nonumber\\ && A^{(1)} \to \tilde A^{(1)} =
dx^\mu\; {\cal B}_\mu (x, \theta, \bar\theta) + d \theta \;\bar
{\cal F} (x, \theta,\bar\theta) + d \bar\theta \;{\cal F} (x,
\theta, \bar\theta),
\end{array} \eqno(3.1)
$$ where the mapping from the 4D local fields to the superfields
are: $A_\mu (x) \to {\cal B}_\mu (x, \theta, \bar\theta)$, $C (x)
\to {\cal F} (x,\theta,\bar\theta)$ and $\bar C (x) \to \bar {\cal
F} (x, \theta, \bar\theta)$. The super-expansion of the
superfields, in terms of the basic fields as well as the secondary
fields, are (see, e.g., [4-7, 10-12]): $$
\begin{array}{lcl}
{\cal B}_\mu (x, \theta, \bar\theta) &=& A_\mu (x) + \theta\; \bar
R_\mu (x) + \bar\theta\; R_\mu (x) + i\; \theta\;\bar\theta \;
S_\mu (x), \nonumber\\ {\cal F} (x, \theta, \bar\theta) & = & C
(x) + i\; \theta\; \bar B_1 (x) + i\; \bar \theta\; B_1 (x) +
i\;\theta\;\bar\theta\; s (x),  \nonumber\\ \bar {\cal F} (x,
\theta, \bar\theta) & = & \bar C (x) + i\; \theta\; \bar B_2 (x) +
i\; \bar \theta\; B_2 (x) + i\;\theta\;\bar\theta \;\bar s (x).
\end{array} \eqno(3.2)
$$ It can be readily seen that, in the limiting case of $(\theta,
\bar\theta) \to 0$, we get back our 4D basic fields ($A_\mu, C,
\bar C$). Furthermore, on the r.h.s. of the above super expansion,
the bosonic (i.e. $A_\mu, S_\mu, B_1, \bar B_1, B_2, \bar B_2$)
and the fermionic ($R_\mu, \bar R_\mu, C, \bar C, s , \bar s$)
fields do match.

At this juncture, we have to recall our observations after
equation (2.2). The nilpotent (anti-)BRST symmetry transformations
basically owe their origin to the cohomological operator $d$. This
is capitalized in the horizontality condition where we impose the
restriction $\tilde d \tilde A^{(1)} = d A^{(1)}$ on the super 1-form
connection $\tilde A^{(1)}$ that contains the superfields defined
on the (4, 2)-dimensional supermanifold. The latter condition
yields the following relationships (see, e.g., for details, in our
earlier works [21-25]): $$
\begin{array}{lcl}
B_1 = \bar B_2 = s = \bar s = 0,\; \qquad  \bar B_1 + B_2 = 0,
\end{array} \eqno(3.3)
$$ where we are free to choose the secondary fields $(B_2, \bar B_1)$
(i.e. $B_2 = B \Rightarrow \bar B_1 = - B$) in terms of the
Nakanishi-Lautrup auxiliary field $B$ of the BRST
invariant Lagrangian density (2.1). The other relations, that
emerge from the above HC (i.e. $\tilde d \tilde A^{(1)} = d
A^{(1)}$), are $$
\begin{array}{lcl}
R_\mu = \partial_\mu C, \qquad \bar R_\mu = \partial_\mu \bar C,
\qquad S_\mu = \partial_\mu B, \qquad 
\partial_\mu {\cal B}_\nu - \partial_\nu {\cal B}_\mu =
\partial_\mu A_\nu - \partial_\nu A_\mu.
\end{array} \eqno(3.4)
$$ At this stage, the super-curvature tensor $\tilde F_{\mu\nu} =
\partial_\mu {\cal B}_\nu - \partial_\nu {\cal B}_\mu$ is {\it
not} equal to the ordinary curvature tensor $F_{\mu\nu} =
\partial_\mu A_\nu - \partial_\nu A_\mu$ as the former contains
Grassmannian dependent terms.

The substitution of the above values (cf. (3.3),(3.4)) of the
secondary fields, in terms of the basic and auxiliary fields of the
Lagrangian density (2.1), leads to $$
\begin{array}{lcl}
{\cal B}^{(h)}_\mu (x, \theta, \bar\theta) &=& A_\mu + \theta\;
\partial_\mu \bar C + \bar\theta\; \partial_\mu C
+ i\; \theta\;\bar\theta \; \partial_\mu B, \nonumber\\ {\cal
F}^{(h)} (x, \theta, \bar\theta) & = & C - i\; \theta\; B,\; \qquad\;
\bar {\cal F}^{(h)} (x, \theta, \bar\theta)  =  \bar C + i\; \bar
\theta\; B,
\end{array} \eqno(3.5)
$$ where the superscript $(h)$ has been used to denote that the
above expansions have been obtained after the application of the
HC. It can be seen that, due to (3.5), we get$$
\begin{array}{lcl}
\partial_\mu {\cal B}^{(h)}_\nu - \partial_\nu {\cal B}^{(h)}_\mu =
\partial_\mu A_\nu - \partial_\nu A_\mu,
\end{array} \eqno(3.6)
$$ where there is {\it no} Grassmannian $\theta$ and $\bar\theta$
dependence on the l.h.s.

In the language of the geometry on the (4, 2)-dimensional
supermanifold, the expansions (3.5) imply that the (anti-)BRST
symmetry transformations $s_{(a)b}$ (and their corresponding
generators $Q_{(a)b}$) for the 4D local fields (cf. (2.7)) are
connected with the translational generators
$(\partial/\partial\theta,
\partial/\partial\bar\theta)$ because the translation of the
corresponding (4, 2)-dimensional superfields, along the
Grassmannian directions of the supermanifold, produces it. Thus,
the Grassmannian {\it independence} of the super curvature tensor
$\tilde F_{\mu\nu}^{(h)} = \partial_\mu {\cal B}_\nu^{(h)} -
\partial_\nu {\cal B}_\mu^{(h)}$ implies that the 4D curvature
tensor $F_{\mu\nu}$ is an (anti-)BRST (i.e. gauge) invariant
physical quantity.

In terms of the superfields, equations (2.3) can be expressed as
$$
\begin{array}{lcl}
&& \mbox{Lim}_{\theta \to 0} \;{\displaystyle
\frac{\partial}{\partial \bar \theta} \Bigl [\; - i\; \bar {\cal
F}^{(h)}\; \bigl \{ \;(\partial^\mu {\cal B}^{(h)}_\mu +
\frac{1}{2} \; B) \;\bigr \}\; \Bigr ]}, \nonumber\\
 && \mbox{Lim}_{\bar\theta \to 0}\;
{\displaystyle \frac{\partial}{\partial \theta} \Bigl [\; +\; i
{\cal F}^{(h)} \;\bigl \{ \;(\partial^\mu {\cal B}^{(h)}_\mu +
\frac{1}{2} \; B \bigr ) \; \}\;\Bigr ]},  \nonumber\\ &&
{\displaystyle \frac{\partial}{\partial \bar\theta} \;
\frac{\partial}{\partial \theta}\; \Bigl [\; \frac{i}{2} \;{\cal
B}^{\mu (h)} {\cal B}^{(h)}_\mu \;+ \;\frac{1}{2}\; {\cal
F}^{(h)}\; \bar {\cal F}^{(h)} \;\Bigr ]} .
\end{array} \eqno(3.7)
$$ These equations are {\it unique} because there is no other way
to express the above equations in terms of the derivatives w.r.t.
Grassmannian variables $\theta$ and $\bar\theta$. Thus, besides
(2.3), there is no other possibility to express the gauge-fixing
and the Faddeev-Popov ghost terms in the language of the off-shell
nilpotent (anti-)BRST symmetry transformations (2.2). The
superfield approach to BRST formulation, therefore, establishes
the uniqueness of (2.3).

To express (2.6) in terms of the
superfields, one has to substitute $B = - (\partial_\mu A^\mu)$ in
(3.5). Thus, the expansion (3.5), in terms of the transformations
(2.4), becomes\footnote{The on-shell nilpotent (anti-)BRST
symmetry transformations $\tilde s_{(a)b}$ can also be obtained by
invoking the (anti-)chiral superfields on the appropriately chosen
supermanifolds (see, e.g. [23] for details).}
 $$
\begin{array}{lcl}
{\cal B}^{(h)}_{\mu (o)} (x, \theta, \bar\theta) &=& A_\mu +
\theta\;
\partial_\mu \bar C + \bar\theta\; \partial_\mu C
- i\; \theta\;\bar\theta \; \partial_\mu (\partial^\rho A_\rho),
\nonumber\\ &\equiv& A_\mu + \theta\; (\tilde s_{ab} A_\mu) +
\bar\theta\; (\tilde s_b A_\mu) + \theta\; \bar \theta (\tilde s_b
\tilde s_{ab} A_\mu), \nonumber\\ {\cal F}^{(h)}_{(o)} (x, \theta,
\bar\theta) & = & C + i\; \theta\; (\partial_\mu A^\mu) \equiv C +
\theta \; (\tilde s_{ab} C), \nonumber\\ \bar {\cal F}^{(h)}_{(o)}
(x, \theta, \bar\theta) &=& \bar C - i\; \bar \theta\;
(\partial_\mu A^\mu) \equiv \bar C + \bar\theta\; (\tilde s_b \bar
C).
\end{array} \eqno(3.8)
$$ We note that (3.5) and (3.8) are the super expansions (after
the application of the HC) which lead to the derivation of the
off-shell nilpotent (anti-)BRST symmetry transformations
$s_{(a)b}$ as well as the on-shell nilpotent (anti-)BRST symmetry
transformations $\tilde s_{(a)b}$, respectively, for the basic
fields $A_\mu, C$ and $\bar C$ of the theory.

The gauge-fixing and
Faddeev-Popov ghost terms of the Lagrangian density (2.5) can also be
expressed in terms of the superfields (3.8). In other
words, ({\it vis-{\`a}-vis} (3.7)), we have the following
equations that are the analogue of (2.6), namely; $$
\begin{array}{lcl}
&& \mbox{Lim}_{\theta \to 0} \;{\displaystyle
\frac{\partial}{\partial \bar \theta} \Bigl [\; + \frac{i}{2}\;
\bar {\cal F}_{(o)}^{(h)}\;(\partial^\mu A_\mu) + i\; {\cal
B}_{\mu(o)}^{(h)} \;\partial^\mu \bar {\cal F}_{(o)}^{(h)})\;
\Bigr ]},  \nonumber\\
 && \mbox{Lim}_{\bar\theta \to 0}\;
{\displaystyle \frac{\partial}{\partial \theta} \Bigl [\; -\;
\frac{i}{2}\; {\cal F}_{(o)}^{(h)} \;(\partial^\mu A_\mu) - i\
{\cal B}_{\mu(o)}^{(h)} \;\partial^\mu  {\cal F}_{(o)}^{(h)})\;
\Bigr ]},  \nonumber\\ &&
{\displaystyle \frac{\partial}{\partial \bar\theta} \;
\frac{\partial}{\partial \theta}\; \Bigl [\; \frac{i}{2} \;{\cal
B}_{(o)}^{\mu (h)} {\cal B}_{\mu(o)}^{(h)} \;+ \;\frac{1}{2}\;
{\cal F}_{(o)}^{(h)}\; \bar {\cal F}_{(o)}^{(h)} \;\Bigr ] }.
\end{array} \eqno(3.9)
$$ 
We know that, for all practical computational purposes, it is
essential to take into account $\tilde s_{(a)b} (\partial_\mu
A^\mu) = 0$ because of the on-shell conditions $\Box C = \Box \bar
C = 0$. The logical reason behind such a restriction (i.e. $\tilde
s_{(a)b} (\partial_\mu A^\mu) = 0$) in (2.6) is encoded in the superfield
approach to BRST formalism as can be seen from a close look at (3.9).

The Lagrangian density (2.1) can be expressed, in terms of the
(4, 2)-dimensional superfields, in the
following distinct and different forms $$
\begin{array}{lcl}
\tilde {\cal L}^{(1)}_B = {\displaystyle - \frac{1}{4}} \tilde
F^{(h)}_{\mu\nu} \tilde F^{\mu\nu (h)} + \mbox{Lim}_{\theta \to
0}\; {\displaystyle \frac{\partial}{\partial\bar\theta}\; \Bigl [
- i \; \bar {\cal F}^{(h)} \bigl (\partial^\mu {\cal B}^{(h)}_\mu
+ \frac{1}{2} \; B \bigr ) \Bigr ]},
\end{array} \eqno(3.10)
$$
$$
\begin{array}{lcl}
\tilde {\cal L}^{(2)}_B = {\displaystyle - \frac{1}{4}} \tilde
F^{(h)}_{\mu\nu} \tilde F^{\mu\nu (h)} +
 \mbox{Lim}_{\bar\theta \to 0}\;
{\displaystyle \frac{\partial}{\partial\theta}\; \Bigl [+ i \;
{\cal F}^{(h)} \bigl (\partial^\mu {\cal B}^{(h)}_\mu +
\frac{1}{2} \; B \bigr ) \Bigr ]},
\end{array} \eqno(3.11)
$$
$$
\begin{array}{lcl}
\tilde {\cal L}^{(3)}_B = - {\displaystyle \frac{1}{4}} \tilde
F^{(h)}_{\mu\nu} \tilde F^{\mu\nu (h)} + {\displaystyle
\frac{\partial}{\partial \bar\theta} \; \frac{\partial}{\partial
\theta}\; \Bigl [+ \frac{i}{2} {\cal B}^{\mu (h)} {\cal
B}^{(h)}_\mu + \frac{1}{2} {\cal F}^{(h)} \bar {\cal F}^{(h)}
\Bigr ]}.
\end{array} \eqno(3.12)
$$ It would be noted that the kinetic energy term $- (1/4) \tilde
F^{(h)}_{\mu\nu} \tilde F^{\mu\nu (h)}$ is independent of the
variables $\theta$ and $\bar\theta$ 
because $\tilde F_{\mu\nu}^{(h)} =
F_{\mu\nu}$. In exactly similar fashion, the Lagrangian density of
(2.5) can be expressed, with the help of the super expansion
(3.8), as $$
\begin{array}{lcl}
\tilde {\cal L}^{(1)}_b = {\displaystyle - \frac{1}{4}} \tilde
F^{(h)}_{\mu\nu (o)} \tilde F^{\mu\nu (h)}_{(o)} + \mbox{Lim}_{\theta \to
0}\;{\displaystyle \frac{\partial}{\partial \bar \theta} \Bigl [\;
+ \frac{i}{2}\; \bar {\cal F}_{(o)}^{(h)}\;(\partial^\mu A_\mu) +
i\; {\cal B}_{\mu(o)}^{(h)} \;\partial^\mu \bar {\cal
F}_{(o)}^{(h)})\; \Bigr ]},
\end{array} \eqno(3.13)
$$
$$
\begin{array}{lcl}
\tilde {\cal L}^{(2)}_b = {\displaystyle - \frac{1}{4}} \tilde
F^{(h)}_{\mu\nu (o)} \tilde F^{\mu\nu (h)}_{(o)} +
 \mbox{Lim}_{\bar\theta \to 0}\;
{\displaystyle \frac{\partial}{\partial \theta} \Bigl [\; -\;
\frac{i}{2}\; {\cal F}_{(o)}^{(h)} \;(\partial^\mu A_\mu) - i\
{\cal B}_{\mu(0)}^{(h)} \;\partial^\mu  {\cal F}_{(o)}^{(h)})\;
\Bigr ]},
\end{array} \eqno(3.14)
$$
$$
\begin{array}{lcl}
\tilde {\cal L}^{(3)}_b = - {\displaystyle \frac{1}{4}} \tilde
F^{(h)}_{\mu\nu (o)} \tilde F^{\mu\nu (h)}_{(o)} + {\displaystyle
\frac{\partial}{\partial \bar\theta} \; \frac{\partial}{\partial
\theta}\; \Bigl [ \;+ \frac{i}{2} {\cal B}_{(o)}^{\mu (h)} {\cal
B}^{(h)}_{\mu(o)} + \frac{1}{2} {\cal F}_{(o)}^{(h)} \bar {\cal
F}_{(o)}^{(h)} \;\Bigr ]}.
\end{array} \eqno(3.15)
$$ The form of the Lagrangian densities (e.g. from (3.10) to
(3.15)) simplify the proof for the (anti-)BRST invariance of the
Lagrangian densities in (2.1) and (2.5).

In the above forms (e.g. from (3.10) to (3.12)) of the Lagrangian
density, the BRST invariance $s_b {\cal L}_B = 0$ and the
anti-BRST invariance $s_{ab} {\cal L}_B = 0$ become very
transparent and simple because the following equalities and
mappings exist, namely; $$
\begin{array}{lcl}
s_b {\cal L}^{(a)}_B = 0 \Rightarrow \mbox{Lim}_{\theta \to 0}
{\displaystyle \frac{\partial}{\partial\bar\theta} \tilde {\cal
L}^{(1)}_B = 0,\; \quad \;s_b \Leftrightarrow \mbox {Lim}_{\theta \to
0} \frac{\partial}{\partial\bar\theta}},\; \quad \;s_b^2 = 0
 \Leftrightarrow \Bigl ({\displaystyle
 \frac{\partial}{\partial\bar\theta}} \Bigr )^2 = 0,
\end{array}\eqno(3.16)
$$
$$
\begin{array}{lcl}
s_{ab} {\cal L}^{(a)}_B = 0 \Rightarrow \mbox{Lim}_{\bar \theta
\to 0} {\displaystyle \frac{\partial}{\partial\theta} \tilde {\cal
L}^{(2)}_B = 0, \quad s_{ab} \Leftrightarrow \mbox
{Lim}_{\bar\theta \to 0}
 \frac{\partial}{\partial\theta}},\; \quad\;
 s_{ab}^2 = 0 \Leftrightarrow \Bigl ({\displaystyle
 \frac{\partial}{\partial\theta}} \Bigr )^2 = 0.
\end{array}\eqno(3.17)
$$ Similarly, the most beautiful relation (3.12), leads to the
(anti-)BRST invariance {\it together}. Here one has to use the
anticommutativity property $s_b s_{ab} + s_{ab} s_b = 0$ in the
language of the translational generators (i.e. $(\partial/\partial
\bar\theta), (\partial/\partial\theta))$ along the Grassmannian
directions of the supermanifold, for its proof. This statement can
be mathematically expressed as $$
\begin{array}{lcl}
s_{(a)b} {\cal L}^{(a)}_B = 0 \Rightarrow  {\displaystyle
\Bigl (\frac{\partial}{\partial\theta} \Bigr )
\frac{\partial}{\partial \bar\theta}} \tilde {\cal L}^{(3)}_B = 0,\;
\qquad \;s_b s_{ab} + s_{ab} s_b = 0  \Leftrightarrow
 {\displaystyle
 \frac{\partial}{\partial\theta} \frac{\partial}{\partial\bar\theta}
 + \frac{\partial}{\partial\bar\theta}
\frac{\partial}{\partial\theta}} = 0.
\end{array}\eqno(3.18)
$$ In exactly similar fashion, the on-shell nilpotent (anti-)BRST
symmetry invariance (i.e. $\tilde s_{(a)b} {\cal L}^{(a)}_b = 0$)
of the Lagrangian density (2.5) can also be captured in the
language of the superfields if we exploit the expressions (3.13)
to (3.15) for the Lagrangian density. In the latter case, the
on-shell nilpotent (anti-)BRST invariance turns out to be like 
(3.16), (3.17) and (3.18) with the replacements: $s_{(a)b} \to \tilde s_{(a)b}, 
\;{\cal L}^{(a)}_B \to {\cal L}^{(a)}_b, \;\tilde {\cal L}^{(1,2,3)}_B 
\to \tilde {\cal L}^{(1,2,3)}_b$.

Mathematically, the (anti-)BRST invariance of the Lagrangian
density (2.1) is captured in the equations (3.16) to (3.18). In
the language of geometry on the (4, 2)-dimensional supermanifold,
the (anti-)BRST invariance corresponds to the Grassmannian
independence of the supersymmetric versions of the Lagrangian
density (2.1). In other words, the translation of the super
Lagrangian densities (i.e. (3.10) to (3.12)), along the
$(\theta)\bar\theta$ directions of the supermanifold, is {\it
zero}. This observation captures the (anti-)BRST invariance of
(2.1).\\

\noindent {\bf 4 (Anti-)BRST symmetries in non-Abelian theory:
Lagrangian approach}\\

\noindent  We begin with the following BRST-invariant Lagrangian
density, in the Feynman gauge, for the four (3 + 1)-dimensional
non-Abelian 1-form gauge theory\footnote{For the non-Abelian 
1-form gauge
theory, the notations used in the Lie algebraic space are: $
A \cdot B = A^a B^a,\; (A \times B)^a = f^{abc} A^b B^c, \;D_\mu C^a =
\partial_\mu C^a + i f^{abc} A_\mu^b C^c \equiv
\partial_\mu C^a + i (A_\mu \times C)^a, \;F_{\mu\nu} = \partial_\mu
A_\nu - \partial_\nu A_\mu + i A_\mu \times A_\nu,\; A_\mu = A_\mu
\cdot T,\; [T^a, T^b] = f^{abc} T^c$ where the Latin indices $a, b, c
= 1, 2, 3.... N$ are in the $SU(N)$ Lie algebraic space. The
structure constant $f^{abc}$ can be chosen to be totally
antisymmetric for any arbitrary semi simple Lie algebra that
includes $SU(N)$, too (see, e.g., [27]). } (see, e.g. [26,27,9])
$$
\begin{array}{lcl}
{\cal L}^{(n)}_B = {\displaystyle - \frac{1}{4} F^{\mu\nu}\cdot
F_{\mu\nu} + B \cdot (\partial_\mu A^\mu) + \frac{1}{2} B \cdot B
- i
\partial_\mu \bar C \cdot D^\mu C},
\end{array} \eqno(4.1)
$$ where the curvature tensor
($F_{\mu\nu}$) is defined through the 2-form $ F^{(2)(n)} =  d
A^{(1)(n)} + i A^{(1)(n)} \wedge  A^{(1)(n)}$. Here the
non-Abelian 1-form gauge connection is $A^{(1)(n)} = dx^\mu (A_\mu \cdot
T)$ and the exterior derivative is $d = dx^\mu \partial_\mu$. The
Nakanishi-Lautrup auxiliary field $B = B \cdot T$ is required for
the linearization of the gauge-fixing term and the (anti-)ghost 
fields $(\bar C)C$ are essential for the
proof of the unitarity in the theory. The latter
fields are fermionic (i.e. $(C^a)^2 = 0, (\bar C^a)^2 = 0, C^a C^b
+ C^b C^a = 0, C^a \bar C^b + \bar C^b C^a = 0$, etc.) in nature.

The above Lagrangian density respects the following off-shell
nilpotent ($(s_b^{(n)})^2 = 0$) BRST symmetry transformations $s^{(n)}_b$,
namely;
$$
\begin{array}{lcl}
&& s^{(n)}_b A_\mu = D_\mu C,\; \qquad s^{(n)}_b C = - {\displaystyle
\frac{i}{2}} (C \times C),\; \qquad s^{(n)}_b \bar C = i B, \nonumber\\ &&
s^{(n)}_b B = 0,\; \qquad s^{(n)}_b F_{\mu\nu} = i (F_{\mu\nu} \times C).
\end{array}\eqno(4.2)
$$ It will be noted that (i) the curvature tensor $F_{\mu\nu}
\cdot T$ transforms here under the BRST symmetry transformation.
However, it can be checked explicitly that the kinetic energy term
$- (1/4) F_{\mu\nu} \cdot F^{\mu\nu}$ remains invariant under the
BRST symmetry transformations, (ii) the nilpotent {\it anti-BRST} symmetry
transformations corresponding to the above BRST symmetry
transformations (4.2) {\it cannot} be defined for the Lagrangian
density (4.1), and (iii) the on-shell nilpotent version of the
above BRST symmetry transformations is also possible if we
substitute, in the above symmetry transformations, $B = -
(\partial_\mu A^\mu)$. The ensuing on-shell (i.e. $\partial_\mu
D^\mu C = 0$) nilpotent BRST symmetry transformations $\tilde s^{(n)}_b$
are $$
\begin{array}{lcl}
&& \tilde s^{(n)}_b A_\mu = D_\mu C,\; \qquad \tilde s^{(n)}_b C = -
{\displaystyle \frac{i}{2}} (C \times C), \nonumber\\ && \tilde
s^{(n)}_b \bar C = - i (\partial_\mu A^\mu),\; \qquad   
\tilde s^{(n)}_b F_{\mu\nu} = i
(F_{\mu\nu} \times C).
\end{array}\eqno(4.3)
$$ The above on-shell nilpotent transformations leave the following Lagrangian
density $$
\begin{array}{lcl}
{\cal L}^{(n)}_b = {\displaystyle - \frac{1}{4} F^{\mu\nu}\cdot
F_{\mu\nu} - \frac{1}{2} (\partial_\mu A^\mu) \cdot (\partial_\rho
A^\rho) - i
\partial_\mu \bar C \cdot D^\mu C},
\end{array} \eqno(4.4)
$$ quasi-invariant because it transforms to a total derivative.

The gauge-fixing and Faddeev-Popov ghost terms of the Lagrangian
densities (4.1) and (4.4) can be written, modulo a total derivative, as a
BRST-exact quantity in terms of the off-shell and on-shell nilpotent BRST
symmetry transformations (4.2) and (4.3). This statement can be
mathematically expressed as follows $$
\begin{array}{lcl}
s^{(n)}_b \Bigl [ - i\; \bar C \cdot \{ (\partial_\mu A^\mu) +
{\displaystyle \frac{1}{2} B \} \Bigr ] =   B \cdot (\partial_\mu
A^\mu) + \frac{1}{2} B \cdot B - i\; \partial_\mu \bar C \cdot
D^\mu C},
\end{array}\eqno(4.5)
$$  
$$
\begin{array}{lcl}
\tilde s^{(n)}_b \Bigl [ + {\displaystyle \frac{i}{2}\; \bar C \cdot
(\partial_\mu A^\mu) +  i\; A_\mu \cdot \partial^\mu \bar C\;
\Bigr ] = - \frac{1}{2}\; (\partial_\mu A^\mu) \cdot
(\partial_\rho A^\rho) - i\; \partial_\mu \bar C \cdot D^\mu C}.
\end{array}\eqno(4.6)
$$ It will be noted that one has to take into account $\tilde s^{(n)}_b
(\partial_\mu A^\mu) = \partial_\mu D^\mu C = 0$ in the above proof of the 
exactness of the expression in (4.6).

The Lagrangian densities that respect the off-shell nilpotent (i.e.
$(s^{(n)}_{(a)b})^2 = 0$) and anticommuting
($s_b^{(n)} s_{ab}^{(n)} + s_{ab}^{(n)} s_b^{(n)} = 0$)
(anti-)BRST symmetry
transformations are
 $$
\begin{array}{lcl}
{\cal L}^{(1)(n)}_b = {\displaystyle - \frac{1}{4} F^{\mu\nu}\cdot
F_{\mu\nu} + B \cdot (\partial_\mu A^\mu) + \frac{1}{2} (B \cdot B
+ \bar B \cdot \bar B) - i
\partial_\mu \bar C \cdot D^\mu C},
\end{array} \eqno(4.7)
$$ $$
\begin{array}{lcl}
{\cal L}^{(2)(n)}_b = {\displaystyle - \frac{1}{4} F^{\mu\nu}\cdot
F_{\mu\nu} - \bar B \cdot (\partial_\mu A^\mu) + \frac{1}{2} (B
\cdot B + \bar B \cdot \bar B) - i D_\mu \bar C \cdot \partial^\mu
 C}.
\end{array} \eqno(4.8)
$$ Here auxiliary fields $B$ and $\bar B$ satisfy the
Curci-Ferrari condition $B + \bar B = - (C \times \bar C)$
[28,29]. It is also evident, from this relation, that $B \cdot
(\partial_\mu A^\mu) - i
\partial_\mu \bar C \cdot D^\mu C = - \bar B \cdot (\partial_\mu
A^\mu) - i D_\mu \bar C \cdot \partial^\mu C$. Furthermore, it
should be re-emphasized that the Lagrangian densities (4.1) and
(4.4) do not respect the anti-BRST symmetry transformations of any
kind. The BRST and anti-BRST symmetry transformations, for the
above Lagrangian densities, are $$
\begin{array}{lcl}
&& s^{(n)}_b A_\mu = D_\mu C,\; \qquad s^{(n)}_b C = - {\displaystyle
\frac{i}{2}} (C \times C),\; \qquad s^{(n)}_b \bar C = i B, \nonumber\\ &&
s^{(n)}_b B = 0,\; \qquad s^{(n)}_b F_{\mu\nu} = i (F_{\mu\nu} \times C),\; \qquad
s^{(n)}_b \bar B = i (\bar B \times C),
\end{array}\eqno(4.9)
$$ $$
\begin{array}{lcl}
&& s^{(n)}_{ab} A_\mu = D_\mu \bar C,\; \qquad s^{(n)}_{ab} \bar C = -
{\displaystyle \frac{i}{2}} (\bar C \times \bar C),\; \qquad s^{(n)}_{ab}
C = i \bar B, \nonumber\\ && s^{(n)}_{ab} \bar B = 0,\; \qquad s^{(n)}_{ab}
F_{\mu\nu} = i (F_{\mu\nu} \times \bar C),\; \qquad s^{(n)}_{ab} B = i (B
\times \bar C).
\end{array}\eqno(4.10)
$$ The above off-shell nilpotent (anti-)BRST symmetry
transformations leave the Lagrangian densities (4.7) as well as
(4.8) quasi-invariant as they transform to some total derivatives.
The gauge-fixing and Faddeev-Popov ghost terms of the Lagrangian
densities (4.7) and (4.8) can be written, in a symmetrical fashion
with respect to $s^{(n)}_b$ and $s^{(n)}_{ab}$, as $$
\begin{array}{lcl}
s^{(n)}_b s^{(n)}_{ab} \Bigl [ {\displaystyle \frac{i}{2}} A_\mu \cdot A^\mu +
C \cdot \bar C \Bigr ] &=& B \cdot (\partial_\mu A^\mu) +
{\displaystyle \frac{1}{2} (B \cdot B + \bar B \cdot \bar B) - i
\partial_\mu \bar C \cdot D^\mu C}, \nonumber\\
&\equiv& - \bar B \cdot (\partial_\mu A^\mu) + {\displaystyle
\frac{1}{2} (B \cdot B + \bar B \cdot \bar B) - i D_\mu \bar C
\cdot \partial^\mu C}.
\end{array}\eqno(4.11)
$$ This demonstrates the key fact that the above gauge-fixing and
Faddeev-Popov ghost terms are (anti-)BRST invariant {\it together}
because of the nilpotency and anticommutativity of the (anti-)BRST 
symmetry transformations $s_{(a)b}^{(n)}$ that are present in the theory. \\

\noindent {\bf 5 (Anti-)BRST invariance in non-Abelian theory:
superfield approach}\\

\noindent To capture (i) the off-shell as well as the on-shell
nilpotent (anti-)BRST symmetry transformations, and (ii) the
invariance of the Lagrangian densities, in the language of the
superfield approach to BRST formalism, we have to consider the 4D
1-form non-Abelian gauge theory on a (4, 2)-dimensional
supermanifold. As a consequence, we have the following mappings:
$$
\begin{array}{lcl}
d\;\;\; \to \;\;\;\;\tilde d &=& dx^\mu\; \partial_\mu\; + \;d
\theta\;
\partial_\theta\; +\; d \bar\theta\; \partial_{\bar\theta},\; \qquad
\;\;\;\tilde d^2 = 0, \nonumber\\ A^{(1)(n)} \to \tilde A^{(1)(n)}
&=& dx^\mu ({\cal B}_\mu \cdot T) (x, \theta, \bar\theta) + d
\theta (\bar {\cal F} \cdot T) (x, \theta,\bar\theta) + d
\bar\theta ({\cal F} \cdot T) (x, \theta, \bar\theta),
\end{array} \eqno(5.1)
$$ where the (4, 2)-dimensional superfields $({\cal B}_\mu \cdot
T, {\cal F} \cdot T, \bar {\cal F} \cdot T)$ are the
generalizations of the 4D basic local fields $(A_\mu \cdot T, C
\cdot T, \bar C \cdot T)$ of the Lagrangian density (4.1), (4.7)
and (4.8). These superfields can be expanded along the
Grassmannian directions of the supermanifold, in terms of the
basic 4D fields, auxiliary fields and secondary fields as
[4,16,19]$$
\begin{array}{lcl}
({\cal B}_\mu \cdot T) (x, \theta, \bar\theta) &=& (A_\mu \cdot T)
(x) + \theta\; (\bar R_\mu \cdot T) (x) + \bar\theta\; (R_\mu
\cdot T) (x) + i\; \theta\;\bar\theta \; (S_\mu \cdot T) (x),
\nonumber\\ ({\cal F}\cdot T) (x, \theta, \bar\theta) & = & (C
\cdot T) (x) + i\; \theta\; (\bar B_1 \cdot T) (x) + i\; \bar
\theta\; (B_1 \cdot T) (x) + i\;\theta\;\bar\theta\; (s \cdot T)
(x),  \nonumber\\ (\bar {\cal F} \cdot T)  (x, \theta, \bar\theta)
& = & (\bar C \cdot T) (x) + i\; \theta\; (\bar B_2 \cdot T) (x) +
i\; \bar \theta\; (B_2 \cdot T) (x) + i\;\theta\;\bar\theta
\;(\bar s \cdot T) (x).
\end{array} \eqno(5.2)
$$ To determine the exact expressions for the secondary fields, in
terms of the basic and auxiliary fields of the theory, we have to exploit the
HC. The horizontality condition, for the non-Abelian gauge theory
is the requirement of the equality of the Maurer-Cartan equation
on the (super) manifolds. In other words, the covariant reduction
of the super 2-form curvature $\tilde F^{(2)(n)}$ to the ordinary
2-form curvature (i.e. $\tilde d \tilde A^{(1)(n)} +
 i \tilde A^{(1)(n)} \wedge \tilde A^{(1)(n)}
= d A^{(1)(n)} + i A^{(1)(n)} \wedge A^{(1)(n)}$) leads
to the determination of the secondary fields in terms of the basic
and auxiliary 
fields of the theory. The ensuing expansions, in terms of the
basic and auxiliary fields, lead to (i) the derivation of the (anti-)BRST
symmetry transformations for the basic fields of the theory, and
(ii) the geometrical interpretations of the nilpotent (anti-)BRST symmetry
transformations (and their corresponding nilpotent generators) for
the basic fields of the theory as the translations of the
corresponding superfields along the Grassmannian directions of the
(4, 2)-dimensional supermanifold (see, e.g., [16,19]).

With the identifications $B_2 = B$ and $\bar B_1 = \bar B$, the
following relationships emerge after the application of the
horizontality condition \footnote{In the rest of our present text,
we shall be using the short-hand notations for all the fields
e.g.: $A_\mu \cdot T = A_\mu,\; C \cdot T = C,\; B\cdot T = B$, etc.,
for the sake of brevity.} (see, e.g., [16]): $$
\begin{array}{lcl}
&& R_\mu = D_\mu C,\; \qquad \bar R_\mu = D_\mu \bar C,\; \qquad B +
\bar B = - (C \times \bar C),\; \qquad s = i (\bar B \times C),
\nonumber\\ && S_\mu = D_\mu B + D_\mu C \times \bar C \equiv -
D_\mu \bar B - D_\mu \bar C \times C, \nonumber\\ && \bar s = - i
(B \times \bar C),\; \qquad B_1 = - {\displaystyle \frac{1}{2} (C
\times C),\; \qquad \bar B_2 = - \frac{1}{2}} (\bar C \times \bar
C).
\end{array}\eqno(5.3)
$$ The substitution of the above expressions, which are obtained
after the application of the horizontality condition, leads to the
following expansions
 $$
\begin{array}{lcl}
&& {\cal B}^{(h)}_\mu (x, \theta, \bar\theta) = A_\mu + \theta
\;D_\mu \bar C + \bar \theta \;D_\mu C + i\; \theta\; \bar\theta
\;(D_\mu B + D_\mu C \times \bar C), \nonumber\\ && {\cal F}^{(h)}
(x, \theta, \bar\theta) = C + i \;\theta \;\bar B - {\displaystyle
\frac{i}{2}\; \bar\theta\; (C \times C) - \theta \; \bar\theta
\;(\bar B \times C)}, \nonumber\\ && \bar {\cal F}^{(h)} (x,
\theta, \bar\theta) = \bar C - {\displaystyle \frac{i}{2}}\;
\theta\; (\bar C \times \bar C) + i \; \bar\theta\; B +
\theta\;\bar\theta \;(B \times \bar C).
\end{array}\eqno(5.4)
$$ The above expansions (see, e.g., our earlier works [16,19]) can
be expressed in terms of the off-shell nilpotent (anti-)BRST
symmetry transformations (4.9) and (4.10).

With the above expansion at our disposal, the gauge-fixing and
Faddeev-Popov terms of the Lagrangian density (4.1) can be
written, modulo a total ordinary derivative, as $$
\begin{array}{lcl}
\mbox{Lim}_{\theta \to 0} \;{\displaystyle
\frac{\partial}{\partial\bar\theta} \Bigl [ - i \bar {\cal
F}^{(h)} \cdot \partial^\mu {\cal B}^{(h)}_\mu  - \frac{i}{2}\;
\bar {\cal F}^{(h)} \cdot B \Bigr ] = B \cdot (\partial_\mu A^\mu)
+ \frac{1}{2} B \cdot B - i\;\partial_\mu \bar C \cdot D^\mu C}.
\end{array}\eqno(5.5)
$$ Furthermore, it can be seen that, due to the validity and
consequences of the horizontality condition, the super curvature
tensor $\tilde F_{\mu\nu}$ has the following form [16,4]$$
\begin{array}{lcl}
\tilde F^{(h)}_{\mu\nu} = F_{\mu\nu} + i \theta (F_{\mu\nu} \times
\bar C) + i \bar\theta (F_{\mu\nu} \times C) - \theta \;
\bar\theta\; ( F_{\mu\nu} \times B + F_{\mu\nu} \times C \times
\bar C).
\end{array}\eqno(5.6)
$$ It is clear from the above relationship that the kinetic energy
term of the present 4D non-Abelian 1-form gauge theory remains
invariant, namely; $$
\begin{array}{lcl}
- {\displaystyle \frac{1}{4} \tilde F^{(h)}_{\mu\nu} \cdot \tilde
F^{\mu\nu (h)} = - \frac{1}{4} } F_{\mu\nu} \cdot  F^{\mu\nu}.
\end{array}\eqno(5.7)
$$ 
The Grassmannian independence of the l.h.s. of (5.7) has deep meaning
as far as physics is concerned. It implies immediately that the
kinetic energy term of the non-Abelian gauge theory is an
(anti-)BRST (i.e. gauge) invariant {\it physical} quantity.

At this juncture, it is worthwhile to point out that one can also
capture the equation (4.6) in the superfield approach to BRST
formalism where the on-shell nilpotent version of the BRST
symmetry transformations (i.e. $\tilde s^{(n)}_b$) plays an important
role. For this purpose, we have to express the superfield
expansion (5.4) for the on-shell nilpotent BRST symmetry
transformation where one has to exploit the replacement $B = - (\partial_\mu
A^\mu)$. With this substitution, the equation (5.4) for the
superfield expansion becomes $$
\begin{array}{lcl}
&& {\cal B}^{(h)}_{\mu (o)}(x, \theta, \bar\theta) = A_\mu +
\theta \;D_\mu \bar C + \bar \theta \;D_\mu C + i\; \theta\;
\bar\theta \;[- D_\mu (\partial^\rho A_\rho) + D_\mu C \times \bar
C], \nonumber\\ && {\cal F}^{(h)}_{(o)} (x, \theta, \bar\theta) =
C + i \;\theta \;\bar B - {\displaystyle \frac{i}{2}\;
\bar\theta\; (C \times C) - \theta \; \bar\theta \;(\bar B \times
C)}, \nonumber\\ && \bar {\cal F}^{(h)}_{(o)} (x, \theta,
\bar\theta) = \bar C - {\displaystyle \frac{i}{2}}\; \theta\;
(\bar C \times \bar C) - i \; \bar\theta\; (\partial_\mu A^\mu) -
\theta\;\bar\theta \;[(\partial_\mu A^\mu) \times \bar C)].
\end{array}\eqno(5.8)
$$ Now, the equation (4.6) can be expressed in terms of the above
superfields, as: $$
\begin{array}{lcl}
&&\mbox{Lim}_{\theta \to 0} \;{\displaystyle
\frac{\partial}{\partial\bar\theta} \Bigl [ \frac{i}{2} \bar {\cal
F}^{(h)}_{(o)} \cdot (\partial^\mu A_\mu)  + i\; 
{\cal B}^{(h)}_{\mu (o)} \cdot \partial^\mu
\bar {\cal F}^{(h)}_{(o)}  \Bigr ]
= -  \frac{1}{2}\; (\partial_\mu A^\mu) \cdot (\partial_\rho
A^\rho) - i\;\partial_\mu \bar C \cdot D^\mu C}.
\end{array}\eqno(5.9)
$$ Furthermore, it will be noted that the analogue of 
(5.6), for the on-shell nilpotent BRST symmetry
transformation (i.e. $\tilde F^{(h)}_{\mu\nu (o)}$),
 can be obtained by the replacement $B = -
(\partial_\mu A^\mu)$. Once again, the equality (5.7) would remain
intact even if we take into account the on-shell nilpotent BRST
symmetry transformations. Thus, we note that the kinetic energy
term (i.e. $(- (1/4) F^{\mu\nu} \cdot F_{\mu\nu} = 
- (1/4) \tilde F^{\mu\nu (h)}_{(o)} \cdot \tilde F^{(h)}_{\mu\nu (o)}$) of the
non-Abelian gauge theory remains independent of the Grassmannian
variables $\theta$ and $\bar\theta$ after the application of the
HC. This statement is true for the off-shell as well as the on-shell
nilpotent (anti-)BRST symmetry transformations. Physically, it
implies that the kinetic energy term for the gauge field of the
non-Abelian theory is an (anti-)BRST (i.e. gauge) invariant
quantity.

The above key observation helps in expressing the Lagrangian
density (4.1) and (4.4) in terms of the superfields (obtained
after the application of HC), as $$
\begin{array}{lcl}
&& \tilde {\cal L}^{(n)}_B = - {\displaystyle \frac{1}{4}} \;
\tilde F^{(h)}_{\mu\nu} \cdot \tilde F^{\mu\nu (h)} +
\mbox{Lim}_{\theta \to 0} \;{\displaystyle
\frac{\partial}{\partial\bar\theta} \Bigl [ - i \bar {\cal
F}^{(h)} \cdot \partial^\mu {\cal B}^{(h)}_\mu  - \frac{i}{2}\;
\bar {\cal F}^{(h)} \cdot B \Bigr ]}, \nonumber\\ && \tilde {\cal
L}^{(n)}_b = - {\displaystyle \frac{1}{4}} \; \tilde
F^{(h)}_{\mu\nu (o)} \cdot \tilde F^{\mu\nu (h)}_{(o)} + \mbox{Lim}_{\theta
\to 0} \;{\displaystyle \frac{\partial}{\partial\bar\theta} \Bigl
[ \frac{i}{2} \bar {\cal F}^{(h)}_{(o)} \cdot (\partial^\mu A_\mu)
+ i\; {\cal B}^{(h)}_{\mu (o)} \cdot \partial^\mu
\bar {\cal F}^{(h)}_{(o)} \;\Bigr ]}.
\end{array}\eqno(5.10)
$$ This result, in turn, simplifies the BRST invariance of the
above Lagrangian density (4.1) and (4.4) (describing the 4D 1-form
non-Abelian gauge theory) as follows $$
\begin{array}{lcl}
\mbox{Lim}_{\theta \to 0} {\displaystyle
\frac{\partial}{\partial\bar\theta}} \tilde {\cal L}^{(n)}_B = 0
\Rightarrow s^{(n)}_b {\cal L}^{(n)}_B = 0,\; \qquad \mbox{Lim}_{\theta
\to 0} {\displaystyle \frac{\partial}{\partial\bar\theta}} \tilde
{\cal L}^{(n)}_b = 0 \Rightarrow \tilde s^{(n)}_b {\cal L}^{(n)}_b = 0.
\end{array}\eqno(5.11)
$$ This is a great simplification because the
total super Lagrangian densities (5.10) remain independent 
of the Grassmannian variable
$\bar\theta$. This key result is encoded in the mapping
$(s^{(n)}_b, \tilde s^{(n)}_b) \Leftrightarrow \mbox{Lim}_{\theta \to 0}
(\partial /\partial\bar\theta)$ and the nilpotency  
$ (s_b^{(n)})^2 = 0, (\tilde s_b^{(n)})^2 = 0, (\partial/\partial \bar\theta)^2 = 0$.

It can be readily checked that the analogues of (5.5) and (5.9)
cannot be expressed as the
derivative w.r.t. the Grassmannian variable $\theta$. To check
this, one has to exploit the super expansions (5.4) and (5.8)
obtained after the application of the HC (in the context of the
derivation of the off-shell as well as the on-shell nilpotent BRST
symmetry transformations $s^{(n)}_{b}$ and $\tilde s^{(n)}_{b}$). It can be
clearly seen that the operation of the 
derivative w.r.t. the Grassmannian variable
$\theta$, on {\it any} combination of the superfields from the
expansions (5.4) and (5.8), does {\it not} lead to the derivation
of the r.h.s. of (5.5) and (5.9). In the language of
the superfield approach to BRST formalism, this is the reason
behind the non-existence of the anti-BRST symmetry transformations
for the Lagrangian densities (4.1) and (4.4).

The form of the gauge-fixing and Faddeev-Popov terms (4.11),
expressed in terms of the BRST and anti-BRST symmetry
transformations {\it together}, can be represented in the language
of the superfields (obtained after the application of HC), as
 $$
\begin{array}{lcl}
{\displaystyle \frac{\partial}{\partial\bar\theta}\;
\frac{\partial}{\partial\theta} \Bigl [ \;\frac{i}{2} {\cal
B}^{(h)}_\mu \cdot {\cal B}^{\mu (h)} + {\cal F}^{(h)} \cdot \bar
{\cal F}^{(h)} \;\Bigr ]} &=& B \cdot (\partial_\mu A^\mu) +
{\displaystyle \frac{1}{2} (B \cdot B + \bar B \cdot \bar B)}
- i \partial_\mu \bar C \cdot D^\mu C.
\end{array}\eqno(5.12)
$$ As a consequence of the above expression, the Lagrangian
densities (4.7) (as well as (4.8)) can be presented, in terms of
the superfields, as $$
\begin{array}{lcl}
\tilde {\cal L}^{(1,2)(n)}_b = - {\displaystyle \frac{1}{4} \tilde
F^{\mu\nu (h)} \cdot \tilde F^{(h)}_{\mu\nu} +
\frac{\partial}{\partial\bar\theta}\;
\frac{\partial}{\partial\theta} \; \Bigl [ \frac{i}{2} {\cal
B}^{(h)}_\mu \cdot {\cal B}^{\mu (h)} + {\cal F}^{(h)} \cdot \bar
{\cal F}^{(h)} \Bigr ]}.
\end{array}\eqno(5.13)
$$ The BRST and anti-BRST invariance of the above super Lagrangian
density (and that of the ordinary 4D Lagrangian densities (4.7)
and (4.8)) is encoded in the following simple equations that are
expressed in terms of the translational generators along the
Grassmannian directions of the (4, 2)-dimensional supermanifold,
namely; $$
\begin{array}{lcl}
\mbox{Lim}_{\theta \to 0} {\displaystyle
\frac{\partial}{\partial\bar\theta}} \tilde {\cal L}^{(1,2)(n)}_b = 0
\Rightarrow s^{(n)}_b {\cal L}^{(1)(n)}_b = 0, \;
\mbox{Lim}_{\bar\theta \to 0} {\displaystyle
\frac{\partial}{\partial\theta}} \tilde {\cal L}^{(1,2)(n)}_b = 0
\Rightarrow s^{(n)}_{ab} {\cal L}^{(2)(n)}_b = 0.
\end{array}\eqno(5.14)
$$ This is a tremendous simplification of the (anti-)BRST invariance of the 
Lagrangian densities (4.7) and (4.8) in the language of the
superfield approach to BRST formalism. In other words, if one is able to show the
Grassmannian independence of the super Lagrangian densities of the
theory, the (anti-)BRST invariance of the 4D theory follows automatically.

In the
language of the geometry on the supermanifold, the (anti-)BRST
invariance of a 4D Lagrangian density is equivalent to the
statement that the translation of the {\it super} version of the
above Lagrangian density, along the Grassmannian directions of
the (4, 2)-dimensional supermanifold, is {\it zero}. Thus, the super 
Lagrangian density of an (anti-)BRST invariant 4D theory is a Lorentz 
scalar, constructed with the help of (4, 2)-dimensional 
superfields (obtained after the 
application of HC), such that, when the partial derivatives w.r.t.
the Grassmannian variables ($\theta$ and $\bar\theta$)
operate on it, the result is zero.

The nilpotency and anticommutativity properties (that are
associated with the conserved (anti-)BRST charges 
and (anti-)BRST symmetry transformations) are
found to be captured very naturally (cf. (3.16)-(3.18)) when we
consider the superfield formulation of the (anti-)BRST invariance
of the Lagrangian density of a given 1-form gauge theory. 
We mention, in passing, that
one could also derive the analogue of the equations (3.16), (3.17)
and (3.18) for the 4D non-Abelian 1-form gauge theory in a
straightforward manner.\\

\noindent {\bf 6 Conclusions}\\

\noindent In our present investigation, we have concentrated
mainly on the (anti-)BRST invariance of the Lagrangian
densities of the free 4D (non-)Abelian 1-form gauge theories (having
no interaction with matter fields) within the framework of the
superfield approach to BRST formalism. We have been able to 
provide the geometrical basis for
the existence of the  
(anti-)BRST invariance in the above 4D theories. 
To be more specific,
we have been able to show that the Grassmannian independence of
the (4, 2)-dimensional super Lagrangian density, expressed in terms of 
the appropriate superfields, is a clear-cut proof that there is an (anti-)BRST
invariance (cf. (3.16), (3.17), (3.18), (5.11),
(5.14)) in the 4D theory.

If the super Lagrangian density could be
expressed as a sum of (i) a Grassmannian independent term, and
(ii) a derivative w.r.t. the Grassmannian variable, then, the
corresponding 4D Lagrangian density will automatically respect
BRST and/or anti-BRST invariance. In the latter piece of the above super
Lagrangian density, the derivative could be {\it either} w.r.t.
$\theta$ {\it or} w.r.t. $\bar\theta$ {\it or} w.r.t. both of them
put {\it together}. More specifically, (i) if the derivative is w.r.t.
$\bar\theta$, the nilpotent symmetry would correspond to the BRST,
(ii) if the derivative is w.r.t. $\theta$, the nilpotent symmetry
would be that of the anti-BRST type, and (iii) if both the derivatives are present
together, both the nilpotent
(anti-)BRST symmetries would be present together 
(and they would turn out to be anticommuting).

For the 4D (non-)Abelian 1-form gauge theories, that are
considered on the (4, 2)-dimensional supermanifold, it is the
HC on the 1-form super connection
$\tilde A^{(1)}$ that plays a
very important role in the derivation of the (anti-)BRST symmetry
transformations. The cohomological origin for the above HC lies in
the (super) exterior derivatives $(\tilde d) d$. This point has
been made quite clear in our discussions after the off-shell as well as
the on-shell nilpotent (anti-)BRST symmetry transformations (2.2),
(2.4), (4.2), (4.3), (4.9) and (4.10). In fact, it is the full
kinetic energy term of the above theories
(owing its origin to the cohomological operator $d = dx^\mu
\partial_\mu$) that remains invariant under the above on-shell 
as well the off-shell nilpotent 
(anti-)BRST symmetry transformations.

The HC produces {\it specifically} the
nilpotent (anti-)BRST symmetry transformations for the gauge and
(anti-)ghost fields because of the fact that the super 1-form
connection $\tilde A^{(1)}/\tilde A^{(1)(n)}$ (cf. (3.1) and
(5.1)) is constructed with a super vector multiplet (${\cal
B}_\mu, {\cal F}, \bar {\cal F}$) which  is the generalization of
the gauge field $A_\mu$ and the (anti-)ghost fields $(\bar C)C$ (of
the ordinary 4D (non-)Abelian 1-form gauge theories) 
to the (4, 2)-dimensional supermanifold. As a
consequence, {\it only} the nilpotent and anticommuting
(anti-)BRST symmetry transformations for the 4D
local fields $A_\mu, C$ and $\bar C$ are obtained when the full
potential of the HC is exploited within the framework of the
above superfield formulation.

It is worthwhile to point out that {\it geometrically} the super
Lagrangian densities, expressed in terms of the (4, 2)-dimensional 
superfields, are equivalent to the sum of the kinetic
energy term and the translations of some composite superfields
(obtained after the application of the HC) along the Grassmannian
directions (i.e. $\theta$ and/or $\bar\theta$) of the (4,
2)-dimensional supermanifold. This observation is distinctly
different from our earlier works on the superfield approach to 2D
(non-)Abelian 1-form gauge theories [24,25,23] which are found to
correspond to
the topological field theories. In fact, for the latter theories,
the total super Lagrangian density turns out to be a total
derivative w.r.t. the Grassmannian variables ($\theta$ and/or
$\bar\theta$). That is to say, even the kinetic energy term of the
latter theories, is able to be expressed as the total derivative
w.r.t. the variables $\theta$ and/or $\bar\theta$.

In our present endeavour, within the framework of the superfield
approach to BRST formalism, we have been able to provide (i) the
logical reason behind the non-existence of the anti-BRST symmetry
transformations for the Lagrangian densities (4.1) and (4.4) for the 4D 
non-Abelian 1-form gauge theory, (ii) the explicit explanation for the
uniqueness of the equations (2.3) and (2.6) for the 4D Abelian 1-form
gauge theory, (iii) the convincing proof for the on-shell
nilpotent (anti-)BRST invariance of the gauge-fixing term (i.e.
$\tilde s_{(a)b} (\partial_\mu A^\mu) = 0, 
\tilde s^{(n)}_{(a)b} (\partial_\mu A^\mu) = 0$) for the 
(non-)Abelian 1-form gauge theories, and (iv) the compelling arguments
for the non-existence of the exact analogue(s) of (2.3) and (2.6) for
the non-Abelian 1-form gauge theory. To the best of our knowledge,
the logical explanations for the above subtle points 
(connected with the 1-form gauge theories) are completely new. 
Thus, the results of our
present work are simple, beautiful and original.

It is worthwhile to mention that our superfield construction and
its ensuing geometrical interpretations are not specific to the 
Feynman gauge (which has been taken into account in our present
endeavor). To corroborate this assertion, we take the simple case
of the 4D Abelian 1-form gauge theory and write the Lagrangian
density (2.1) in the arbitrary gauge 
$$
\begin{array}{lcl}
{\cal L}^{(a, \xi)}_B = {\displaystyle - \frac{1}{4}\; F^{\mu\nu}
F_{\mu\nu} \;+ \;B\; (\partial_\mu A^\mu)\; +\; \frac{\xi}{2}\; B^2
- i\;\partial_\mu \bar C\;\partial^\mu C},
\end{array} \eqno(6.1)
$$
where $\xi$ is the gauge parameter.
It is elementary to check that, in the limit $\xi \to 1$, we get
back our Lagrangian density (2.1) for the Abelian theory in the Feynman gauge.

The analogue of the equation (2.3) (for the gauge-fixing and Faddeev-Popov
ghost terms in the case of the arbitrary gauge) can be expressed as 
$$
\begin{array}{lcl}
&& {\displaystyle s_b \Bigl [\;- i \;\bar C \;\{ (\partial_\mu
A^\mu) + \frac{\xi}{2}\; B\}]}, \;\;\;\qquad\;\;\;
 {\displaystyle s_{ab} \Bigl [\;+ i\;  C\; \{ (\partial_\mu
A^\mu) + \frac{\xi}{2}\; B\} \; \Bigr ]}, \nonumber\\
&& {\displaystyle s_b\; s_{ab} \;\Bigl [\; \frac{i}{2}\; A_\mu\;
A^\mu + \frac{\xi}{2}\; C \;\bar C\; \Bigr ]}.
\end{array} \eqno(6.2)
$$
The above expression can be easily generalized to the analogues of
the equations (3.10)---(3.12) in terms of the superfields by taking
the help of (3.8). Thus, the geometrical interpretations remain intact
even in the case of the arbitrary gauge.

In a similar fashion, for the 
4D non-Abelian 1-form gauge theory, the equations (4.5), (4.6) and (4.11) 
can be generalized to the case of arbitrary gauge and, subsequently, can be
expressed in terms of superfields as the analogues of (5.5), (5.9) and (5.12).
Finally, we can obtain the analogues of (5.7), (5.10) and (5.13) which will
lead to the derivation of the analogues of (5.11) and (5.14). Thus, we
note that geometrical interpretations, in the arbitrary gauge, remain the same 
for the 4D (non-)Abelian 1-form gauge theory within the framework of our
superfield approach to BRST formalism.

Our present work
can be generalized to the case of the interacting 4D 
(non-)Abelian 1-form gauge theories where there exists an explicit coupling
between the gauge field and the matter fields. In fact, our earlier
works [14-18] might turn out to be quite handy in attempting the
above problems. It seems to us that it is the combination of the
HC and the restrictions, owing their origin to the (super)
covariant derivative on the matter (super) fields and their intimate 
connection with the (super) curvatures, that would play a
decisive role in proving the existence of the (anti-)BRST invariance
for the above gauge theories.

It is gratifying to state that 
we have accomplished the above goals in our very recent 
endeavours [30-32]. In fact, we have been able to provide the geometrical basis 
for the existence of the (anti-)BRST invariance, in the context of the interacting 
(non-)Abelian 1-form gauge theories with Dirac as well as complex scalar fields,
within the framework of the augmented superfield approach to BRST formalism.
As it turns out, here too, the super Lagrangian density is found to be 
independent of the Grassmannian variables.

In our earlier works [33-35], we have been able to show the existence
of the nilpotent (anti-)BRST and (anti-)co-BRST symmetry transformations
for the 4D free Abelian  2-form gauge theory. We have also established the 
quasi-topological nature of it in [35]. In a recent work [36], the nilpotent 
(anti-)BRST symmetry transformations have been captured in the framework 
of the superfield formulation. It would be a very nice endeavour to study 
the (anti-)BRST and (anti-)co-BRST invariance of the 4D Abelian 2-form gauge theory
within the framework of superfield formulation. At present, this issue and 
connected problems in the context of the 4D free Abelian 2-form gauge theory are 
under intensive investigation and our results would be reported in our forthcoming 
future publications [37].\\

\noindent
{\bf Acknowledgement:}
Financial support from the
Department of Science and Technology (DST), Government of India,  
under the SERC project sanction grant No: - SR/S2/HEP-23/2006, is gratefully acknowledged.

\end{document}